\documentclass[aps,twocolumn,prl,tightenlines,floatfix,superscriptaddress]{revtex4-2}

\usepackage[breaklinks=true]{hyperref}
\usepackage{graphicx}
\usepackage[english]{babel}
\usepackage{amsmath}
\usepackage{amssymb}
\usepackage{times}
\usepackage{color}

\begin{document}

\title{Spectral study of the pseudogap in unitary Fermi gases}

 \author{Chuping Li}
 \affiliation{Hefei National Research Center for Physical Sciences at the Microscale and School of Physical Sciences, University 
   of Science and Technology of China, Hefei, Anhui 230026, China}
 \affiliation{Shanghai Research Center for Quantum Science and CAS Center for Excellence in Quantum Information and Quantum Physics, 
 University of Science and Technology of China, Shanghai 201315, China}
 \affiliation{Hefei National Laboratory, University of
  Science and Technology of China, Hefei 230088, China}
 \author{Lin Sun}
 \affiliation{Hefei National Laboratory, University of
  Science and Technology of China, Hefei 230088, China}
 \affiliation{Shanghai Research Center for Quantum Science and CAS Center for Excellence in Quantum Information and Quantum Physics, 
 University of Science and Technology of China, Shanghai 201315, China}
 \author{Kaichao Zhang}
 \affiliation{Hefei National Research Center for Physical Sciences at the Microscale and School of Physical Sciences, University 
   of Science and Technology of China, Hefei, Anhui 230026, China}
 \affiliation{Shanghai Research Center for Quantum Science and CAS Center for Excellence in Quantum Information and Quantum Physics, 
 University of Science and Technology of China, Shanghai 201315, China}
 \affiliation{Hefei National Laboratory, University of
  Science and Technology of China, Hefei 230088, China}
 \author{Junru Wu}
 \affiliation{Hefei National Research Center for Physical Sciences at the Microscale and School of Physical Sciences, University 
   of Science and Technology of China, Hefei, Anhui 230026, China}
 \affiliation{Shanghai Research Center for Quantum Science and CAS Center for Excellence in Quantum Information and Quantum Physics, 
 University of Science and Technology of China, Shanghai 201315, China}
 \affiliation{Hefei National Laboratory, University of
  Science and Technology of China, Hefei 230088, China}
 \author{Yuxuan Wu}
 \affiliation{Hefei National Research Center for Physical Sciences at the Microscale and School of Physical Sciences, University 
   of Science and Technology of China, Hefei, Anhui 230026, China}
 \affiliation{Shanghai Research Center for Quantum Science and CAS Center for Excellence in Quantum Information and Quantum Physics, 
 University of Science and Technology of China, Shanghai 201315, China}
 \affiliation{Hefei National Laboratory, University of
  Science and Technology of China, Hefei 230088, China}
 \author{Dingli Yuan}
 \affiliation{Hefei National Research Center for Physical Sciences at the Microscale and School of Physical Sciences, University 
   of Science and Technology of China, Hefei, Anhui 230026, China}
 \affiliation{Shanghai Research Center for Quantum Science and CAS Center for Excellence in Quantum Information and Quantum Physics, 
 University of Science and Technology of China, Shanghai 201315, China}
 \affiliation{Hefei National Laboratory, University of
  Science and Technology of China, Hefei 230088, China}
 \author{Pengyi Chen}
 \affiliation{Hefei National Research Center for Physical Sciences at the Microscale and School of Physical Sciences, University 
   of Science and Technology of China, Hefei, Anhui 230026, China}
 \affiliation{Shanghai Research Center for Quantum Science and CAS Center for Excellence in Quantum Information and Quantum Physics, 
 University of Science and Technology of China, Shanghai 201315, China}
 \affiliation{Hefei National Laboratory, University of
  Science and Technology of China, Hefei 230088, China}
 \author{Qijin Chen}
 \email[Corresponding author: ]{qjc@ustc.edu.cn}
 \affiliation{Hefei National Research Center for Physical Sciences at the Microscale and School of Physical Sciences, University 
   of Science and Technology of China, Hefei, Anhui 230026, China}
 \affiliation{Shanghai Research Center for Quantum Science and CAS Center for Excellence in Quantum Information and Quantum Physics, 
 University of Science and Technology of China, Shanghai 201315, China}
 \affiliation{Hefei National Laboratory, University of
  Science and Technology of China, Hefei 230088, China}

\date{\today}
\begin{abstract}
The recent observation of a pseudogap in unitary Fermi gases [Li et
  al., Nature 626, 288 (2024)] provides strong evidence for a pairing
origin of the pseudogap in Fermi superfluids. Here we present a
quantitative spectral study of these gases, comparing our
theoretically calculated momentum‑resolved rf/microwave spectra with
the experimental data. Using an iterative treatment of the fermion
self‑energy beyond the previous pseudogap approximation, based on a
pairing fluctuation theory that incorporates both particle‑particle
and particle‑hole $T$ matrices with self‑consistent feedback, we
achieve excellent agreement with the measured spectral functions and
extracted pseudogap parameters. Our results not only explain the
experimental data microscopically but also reinforce the pairing
origin for the pseudogap and the pairing‑fluctuation theory of Fermi
superfluidity.
\end{abstract}
\maketitle

Atomic Fermi gases have provided an ideal platform for simulating and
studying the physics of Fermi superfluidity and superconductivity in
solids \cite{chen2005PR}, such as high $T_c$ superconductivity in
cuprates. One central issue that perplexes our understanding of the
high $T_c$ superconductivity is the very existence of a pseudogap in
the single-particle excitation spectrum in the optimally doped and
underdoped cuprates \cite{Timusk1999RPP}. Assuming a pairing origin
for the pseudogap, as advocated in the preformed pair scenario among
competing theories for high $T_c$ superconductivity, one would expect
to see a pseudogap in atomic Fermi gases when the pairing interaction
is tuned to be strong. Indeed, this was proposed about the same time
\cite{Stajic2004PRA} when Fermi superfluidity was realized
experimentally in these gases two decades ago
\cite{Regal2004PRL}

The pairing and superfluid phenomena in ultracold atomic Fermi gases,
particularly in the unitary regime, have been extensively studied over
the past two decades.  A key focus is the pseudogap phenomenon
\cite{chen2005PR,Chen2014FP,Mueller2017RoPP}; the existence of a
pseudogap in strongly interacting Fermi gases will demonstrate that a
strong pairing interaction can lead to the formation of a pseudogap
\cite{chen2005PR}, and thus provide naturally a plausible explanation
for the widespread pseudogap phenomena in high $T_c$ superconductors
\cite{Timusk1999RPP}.  Conclusive evidence came from very recent
high-precision momentum-resolved microwave spectroscopy measurements
in a homogeneous unitary Fermi gas of $^6$Li \cite{li2024nature}.  It
is the purpose of this work to provide microscopic calculations that
are in \emph{quantitative} agreement with the experiment, thereby
strengthening the support for both the pairing origin of the pseudogap
and our particular pairing fluctuation theory \cite{chen1998PRL}.

Pseudogap can be directly probed via measuring the fermion spectral
function using, e.g., angle-resolved photoemission spectroscopy
(ARPES) in solid state superconductors \cite{Ding1996nature} or its
counterpart, momentum-resolved rf or microwave spectroscopy, in atomic
Fermi gases \cite{Stewart2008nature,Chen2009RoPP}. Indeed, the
existence of the pseudogap in Fermi gases has been experimentally
investigated in a trap using either momentum integrated
\cite{Chin2004science} or momentum-resolved rf spectroscopy
\cite{Stewart2008nature}.  However, these earlier rf measurements were
plagued by trap inhomogeneity, limited resolution, low signal-to-noise
ratio, and final-state interactions \cite{Perali2008PRL,He2009PRL},
and thus have allowed room for ambiguity in explaining the rf
spectra, with possible alternative interpretations without invoking
fermion pairing \cite{Schneider2010PRA,Nascimbene2011PRL}.  The
measurements of Li et al \cite{li2024nature} were free of the
complications caused by the trap inhomogeneity and the well-known
final-state effects \cite{Schunck2008nature}.  The evidence for
pseudogap includes clear, simultaneous observation of the two BCS-like
branches of the quasiparticle dispersion in a \emph{homogeneous}
unitary Fermi gas of $^6$Li both below and \emph{above} the superfluid
transition temperature $T_c$, and quantitative measurements of the
excitation gap and Cooper pair decay rate as a function of
temperature.

In this Letter, we consider a homogeneous three-dimensional (3D)
ultracold Fermi gas at unitarity with short-range $s$-wave pairing
interaction $V_{\mathbf{k},\mathbf{k}^{\prime}} = g < 0$, which
provides a good description of the $^6$Li gases studied in
Ref.~\cite{li2024nature}.  We use a pairing fluctuation theory
\cite{Chen2016SR} that incorporates particle-hole (ph) channel effects
in the particle-particle scattering $T$ matrix. To get quantitatively
accurate results of the spectral function, we go beyond previous
approximations for the pseudogap self-energy and treat it with
elaborate iteration with full numerics.  We arrive at a spectral
function $A(\mathbf{k},\omega)$ that is in quantitative agreement with
experimental data, with two BCS-like quasiparticle branches in the 3D
spectral intensity map and the right amount of broadening.  Treating
these theoretically generated data with the same experimental
procedure, we fit the loci of the spectral intensity peaks with
BCS-like dispersions, and the energy distribution curves (EDCs) with
the same phenomenological model for the self-energy. The extracted
pairing (pseudo)gap, Hartree energy $U$, and effective fermion mass
$m^*$, as well as the inverse pair lifetime $\Gamma_0$ and
single-particle scattering rate $\Gamma_1$, are in quantitative
agreement with experiment, both below and above $T_\text{c}$. This
also includes the thermally activated exponential behavior of
$\Gamma_0$ and nearly constant behavior of $\Gamma_1$ versus $T$.  Our
results not only provide a quantitative explanation for the experiment,
but also manifest that our theory captures the right physics.


To be concrete, the fermion self-energy in our
theory \cite{chen2005PR,chen1998PRL} contains two terms, associated
with the contributions of the Cooper pair condensate and
finite-momentum pairing fluctuations, given by $\Sigma_\text{sc}(K) =
-\Delta_\text{sc}^2 G_0(-K) $ and $\Sigma_\text{pg}(K) =
\sum_{Q\neq0}t_\text{pg}(Q)G_{0}(Q-K) $, respectively. Here the
superfluid order parameter $\Delta_\text{sc}$ vanishes at $T\ge T_c$,
and $G_0(K) = 1/(\mathrm{i}\omega - \xi_\mathbf{k}^{})$ is the bare
Green's function, with the free fermion dispersion $\xi_\mathbf{k}^{}
\equiv\epsilon_\mathbf{k}^{}-\mu'= k^2/2m -\mu^\prime$ and chemical
potential $\mu^\prime$, which contains a shift due to the Hartree
energy (See Equation (2.4) of Ref.~\cite{ChenPhD}). Here the
Hartree shift is generalized to include the whole diagonal part of the
self-energy at the Fermi level, so that $G_0^{-1}(0,k_\mu) =
G^{-1}(0,k_\mu) = 0$. The $T$-matrix associated with finite-momentum
pairs, $t_\text{pg}(Q)=1/(g^{-1}+\chi(Q))$, is given by an infinite
series of particle-particle scattering ladder diagrams, with pair
susceptibility $\chi(Q)=\sum_{K} G_0(Q-K)G(K)$, where $G(K)$ is the
full Green's function. Note that this mixed form of $\chi(Q)$, derived
from the equations of motion approach \cite{Kadanoff1961,ChenPhD},
distinguishes this pairing fluctuation theory from other
$T$-matrix-based theories \cite{Levin2010}.
Now the total self-energy is given by
\begin{eqnarray}
	\label{eq:selfenergy}
	\Sigma(K)&=&\Sigma_\text{sc}(K)+\Sigma_\text{pg}(K)\nonumber\\
	&=&-\Delta^{2}_\text{sc}G_{0}(-K)+\sum_{Q\neq0}t_\text{pg}(Q)G_{0}(Q-K)\,.
\end{eqnarray}
Here and throughout, we set the volume to unity, take the natural units
$\hbar=k_\text{B}=1$, and use the four-momentum notation $K\equiv
(\mathbf{k}, \mathrm{i}\omega_n)$, $Q\equiv (\mathbf{q},
\mathrm{i}\Omega_l)$, $\sum_Q\equiv T\sum_l\sum_\mathbf{q}$, etc, 
where $\omega_n$ (fermionic) and $\Omega_l$ (bosonic) are Matsubara
frequencies \cite{chen1998PRL}.

The Thouless criterion for superfluidity gives $t^{-1}_\text{pg}(0) =
0$. This leads to the pseudogap approximation $\Sigma_\text{pg}(K)
\approx -\Delta_\text{pg}^2G_0(-K)$ for $T\le T_c$ and also above
$T_c$, as long as the pair chemical potential $\mu_\text{p}$ is
small. This defines the pseudogap parameter via $\Delta_\text{pg}^2 =
-\sum_Q t_\text{pg}(Q)$. It brings both $\Sigma(K)$ and $G(K)$ into
the simple BCS-like form, with $\Sigma(K)=-\Delta^2G_{0}(-K)$ and
$G(K)={u_{\mathbf{k}}^{2}}/(\mathrm{i}\omega_{n}-E_{\mathbf{k}})
+{v_{\mathbf{k}}^{2}}/(\mathrm{i}\omega_{n}+E_{\mathbf{k}})$, where
the coherence factors are
$u_{\mathbf{k}}^{2}=(1+\xi_{\mathbf{k}}/E_{\mathbf{k}})/2$,
$v_{\mathbf{k}}^{2}=(1-\xi_{\mathbf{k}}/E_{\mathbf{k}})/2$, with
$E_{\mathbf{k}}=\sqrt{\xi_{\mathbf{k}}^{2}+\Delta^{2}}$, and the
pairing gap $\Delta=\sqrt{\Delta^{2}_\text{sc}+\Delta^{2}_\text{pg}}$.
This BCS-like $G(K)$ then yields the spectral function 
\begin{equation}
  \label{eq:iniAkw}
A_\text{init}(\mathbf{k},\omega)=2\pi\left[u_\mathbf{k}^{2}
  \delta(\omega-E_\mathbf{k})+v_\mathbf{k}^{2}\delta(\omega+E_\mathbf{k})\right]\,,
\end{equation}
which will be used as the initial input in our iterative numerical
procedure described below.

While the pseudogap approximation makes it easy to solve for $T_c$ and
the gap parameters, it is an oversimplification. It drops the diagonal
part of the self-energy so that $\Sigma_\text{pg}$ vanishes when
$\Delta_\text{pg}=0$. The resulting spectral function is obviously too
sharp compared with that one would expect from the convoluted form of
$\Sigma_\text{pg}$; The finite-momentum pair distribution will spread
out the spectral weight.

To enable quantitative comparisons with experiment
\cite{li2024nature}, we need to include the contribution of
particle-hole fluctuations as well. Following Ref.~\cite{Chen2016SR},
we incorporate the entire particle-hole $T$ matrix $t_\text{ph}$ such
that it serves as a screened pairing interaction, and causes a
non-uniform shift of the inverse pairing interaction, $1/g$, by the
particle-hole susceptibility $\langle \chi_\text{ph}\rangle $ averaged
near the Fermi surface.  Hence the finite-momentum $T$-matrix becomes
$t_\text{pg}(Q)=1/(g^{-1}+\chi(Q)+\langle \chi_\text{ph}\rangle )$.
With the BCS-like $G(K)$, $t^{-1}_\text{pg}(Q)$ is given by
\begin{eqnarray}
	\label{eq:tmatrix}
	t^{-1}_\text{pg}({\mathbf q},\mathrm{i}\Omega_{n})\!&=&\!\sum_\mathbf{k}\left[\frac{1-f(E_\mathbf{k})-f(\epsilon_\mathbf{k-q})}{E_\mathbf{k}+\epsilon_\mathbf{k-q}-\mathrm{i}\Omega_{n}}u_{\mathbf{k}}^{2}\right.\nonumber\\\!&-&\!\left.\frac{f(E_\mathbf{k})-f(\epsilon_\mathbf{k-q})}{E_\mathbf{k}-\epsilon_\mathbf{k-q}+\mathrm{i}\Omega_{n}}v_{\mathbf{k}}^{2}\right]+\langle \chi_\text{ph}\rangle+\frac{1}{g}\,.
\end{eqnarray}
Furthermore, rather than using the pseudogap approximation, we
calculate $\Sigma_\text{pg}(K)$ directly using the full convolution in
Eq.~(\ref{eq:selfenergy}), which leads to
\begin{equation}
  \label{eq:realpseudo}
	\Sigma_\text{pg}^\text{R}(\mathbf{k},\omega)=-\!\sum_\mathbf{q}\!\int\!{\frac{{\mathrm{d}}\Omega}{\pi}}{\frac{b(\Omega)+f(\xi_{\mathbf{q}-\mathbf{k}})}{\xi_{\mathbf{q}-\mathbf{k}}\!-\Omega+\omega+\mathrm{i}0^{+}}}\,
  \text{Im}\,t_\text{pg}^\text{R}(\mathbf{q},\Omega)\,,
\end{equation}
after analytical continuation.  Here $b(x)$ and $f(x)$ are Bose and
Fermi distribution functions, respectively. The $b(x)$ term describes
a dominantly off-diagonal BCS-like self-energy due to pair formation.
The $f(x)$ term, which is dropped in the pseudogap approximation,
reflects the interaction effect of a fermion with the particle-hole
bubbles from the medium, and has a regular, dominantly ``diagonal''
contribution to the Hartree self-energy.  The imaginary part of the
self-energy thus reflects the finite lifetimes of quasiparticles
associated with each process.
\begin{figure}[h]
	\centering
	\includegraphics[clip,width=2.8in]{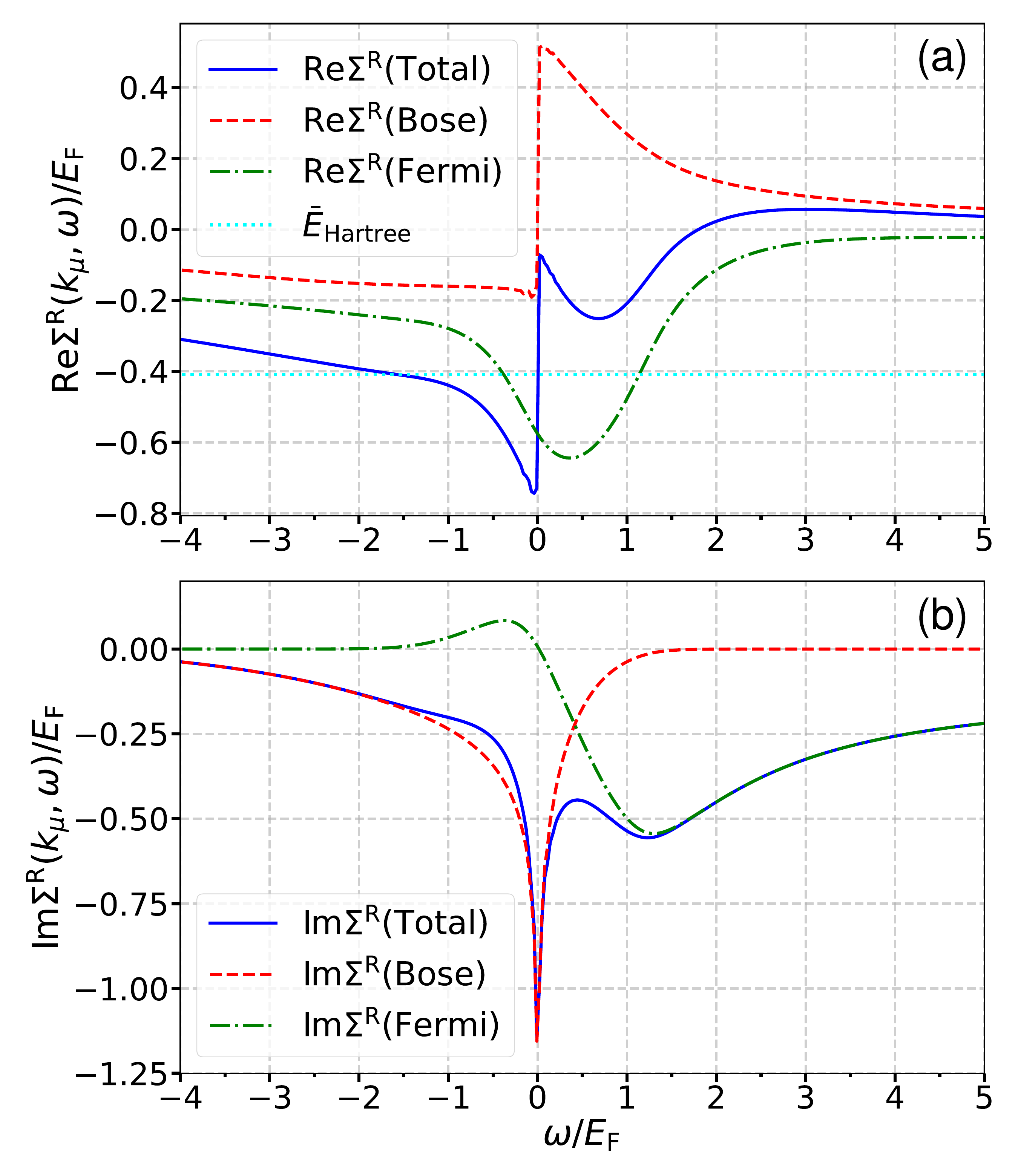}\vspace*{-2ex}
	\caption{ (a) Real and (b) imaginary parts of the retarded
          self-energy $\Sigma^{\text{R}}(k_{\mu},\omega)$ for a
          unitary Fermi gas at $|\mathbf{k}|=k_{\mu}$ and
          $T_{\text{c}}$.  Red dashed and green dash-dotted lines show
          the Bose and Fermi components of total self-energy (blue
          solid lines), respectively, along with the average Hartree
          energy $\bar{E}_{\text{Hartree}}$ (cyan dotted line). }
	\label{fig:Sigma}
\end{figure} 

Using the Kramers-Kronig relation, we obtain the real and imaginary
parts of $\Sigma_\text{pg}^\text{R}(\mathbf{k},\omega)$, as shown in
Fig.~\ref{fig:Sigma}.  Note that the off diagonal self-energy must
necessarily vanish at $k=k_\mu$ and $\omega=0$, where
$k_\mu=\sqrt{2m\mu'}$ corresponds to the wave vector on the Fermi
surface, where the back-bending of the BCS-like quasiparticle
dispersion occurs. This requires that
$\bar{E}_{\text{Hartree}}={\text{Re}}\Sigma^\text{R}_{\text{pg}}(k_\mu,0)$. Thus
the physical chemical potential is given by
$\mu=\mu'+\bar{E}_{\text{Hartree}}$. Note that the main effect of the
Hartree-dominated diagonal self-energy is a chemical potential shift
and slight fermion mass renormalization (It should now be
clear that our bare Green's function $G_0(K)$ contains the diagonal
self-energy via the shifted chemical potential $\mu^\prime$.)  Now
the spectral function is
\begin{equation}
\label{eq:Akw}
A(\mathbf{k},\omega)=\frac{-2\,{\text{Im}}\Sigma^\text{R}(\mathbf{k},\omega)}{[\omega-\epsilon_\mathbf{k}+\mu-\text{Re}\Sigma^{\text{R}}(\mathbf{k},\omega)]^2+[{\text{Im}}\Sigma^{\text{R}}(\mathbf{k},\omega)]^2}\,.
\end{equation}
The pseudogap in the spectral function originates from the negative
peak around $\omega=0$ in
$\text{Im}\Sigma^{\text{R}}(k_{\mu},\omega)$, where $\Sigma^\text{R} =
\Sigma_\text{sc}^\text{R} + \Sigma_\text{pg}^\text{R}$.  Detailed
derivations of our iterative framework and quantitative numerical
results in the BCS and BEC regimes are given in a companion paper
\cite{Regular}.


\begin{figure}
\centering
\includegraphics[clip,width=3.4in]{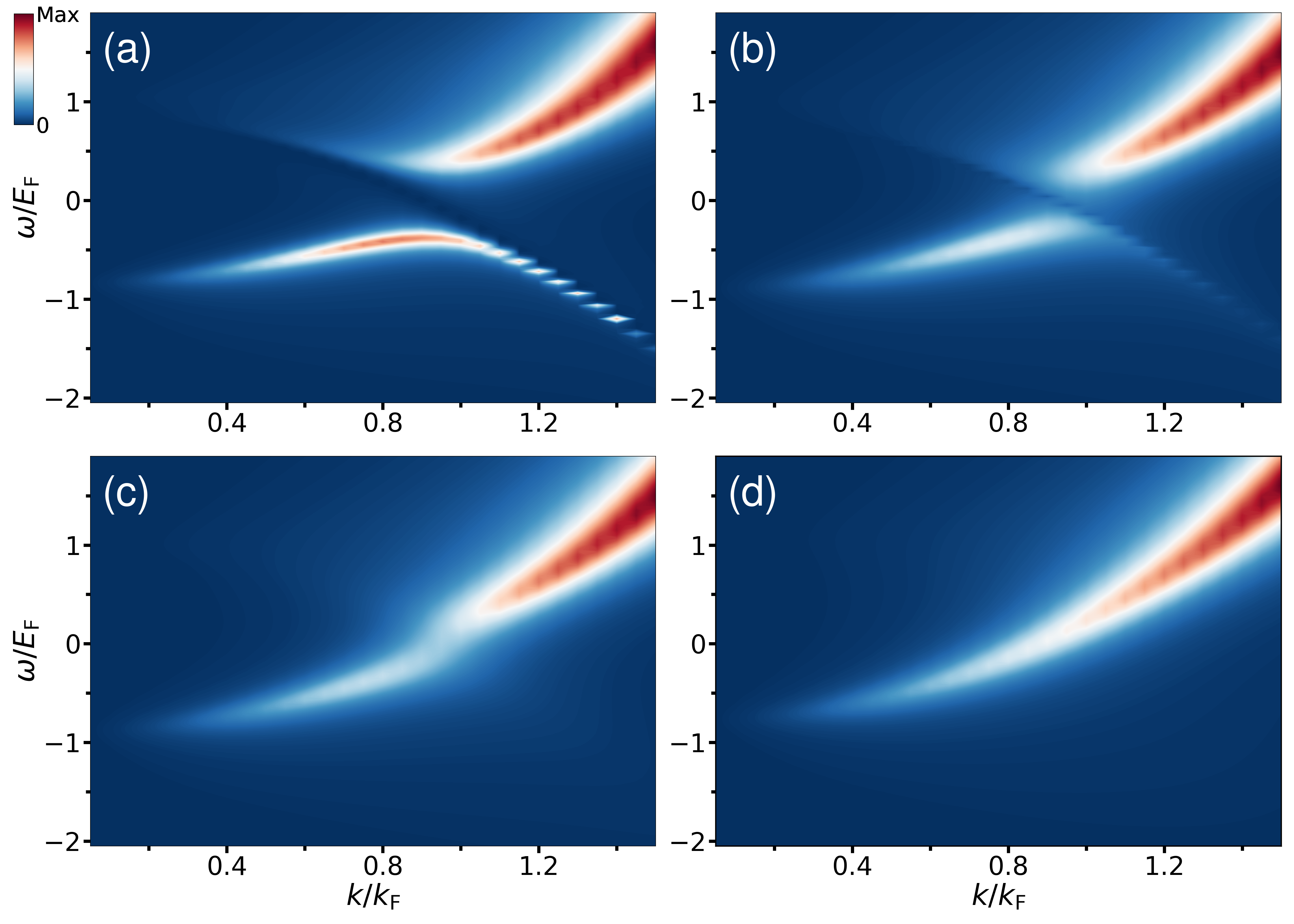}\vspace*{-2ex}
\caption{ Contour plot of $\mathbf{k}^2 A(\mathbf{k},\omega)$ at (a)
  $T/T_\text{c}=0.77$, (b) $1$, (c) $1.11$, and (d) $1.51$, with
  $T_\text{c}/T_\text{F} = 0.2$,  showing quasiparticle dispersions
  evolving from BCS-like gapped branches to a single S-shaped branch
  with a decreasing $\Delta$.  }
\label{fig:Akw1}
\end{figure}

Shown in Fig.~\ref{fig:Akw1} are contour plots of the 
spectral intensity $\mathbf{k}^{2}A(\mathbf{k},\omega)$ as a function
of $k=|\mathbf{k}|$ and $\omega$ at (a) $T/T_\text{c}=0.77$, (b) $1$,
(c) $1.11$ and (d) $1.51$, matching temperatures in
Ref.~\cite{li2024nature}, with $T_\text{c}/T_\text{F} = 0.2$.  At
$T/T_\text{c} \le 1$ in panels (a) and (b), two clearly resolved
excitation branches, with a sizable pairing gap $\Delta$, correspond
to the particle- and hole-like dispersions.  Both branches exhibit a
back-bending behavior around $k_\mu \simeq 0.93k_\text{F}$,
characteristic of BCS-like dispersions.  As temperature rises from
panel (b) to (c), $\Delta$ shrinks, resulting in a hybridization
between the two branches and the emergence of an $S$-shaped
dispersion, as observed in Ref.~\cite{li2024nature}. Intuitively, at
higher $T$ with smaller gap, the spectral weight of the low $k$ part
of the upper branch and the large $k$ part of the lower branch is
expected to become smaller with a wider spread in frequency. This
renders the spectral peak essentially invisible in these two parts, hence
leaving an $S$-shaped dispersion.  This also indicates less
well-defined quasiparticles at this temperature above $T_\text{c}$, so that the
dispersion, as given by the locus of the spectral peak, evolves
continuously from the lower to the upper branch.  At even higher
$T/T_\text{c}=1.51$ in panel (d), $\Delta$ becomes so small that the
$S$ shape is invisible and the dispersion looks simply parabolic.

\begin{figure}
\centering
\includegraphics[clip,width=3.45in]{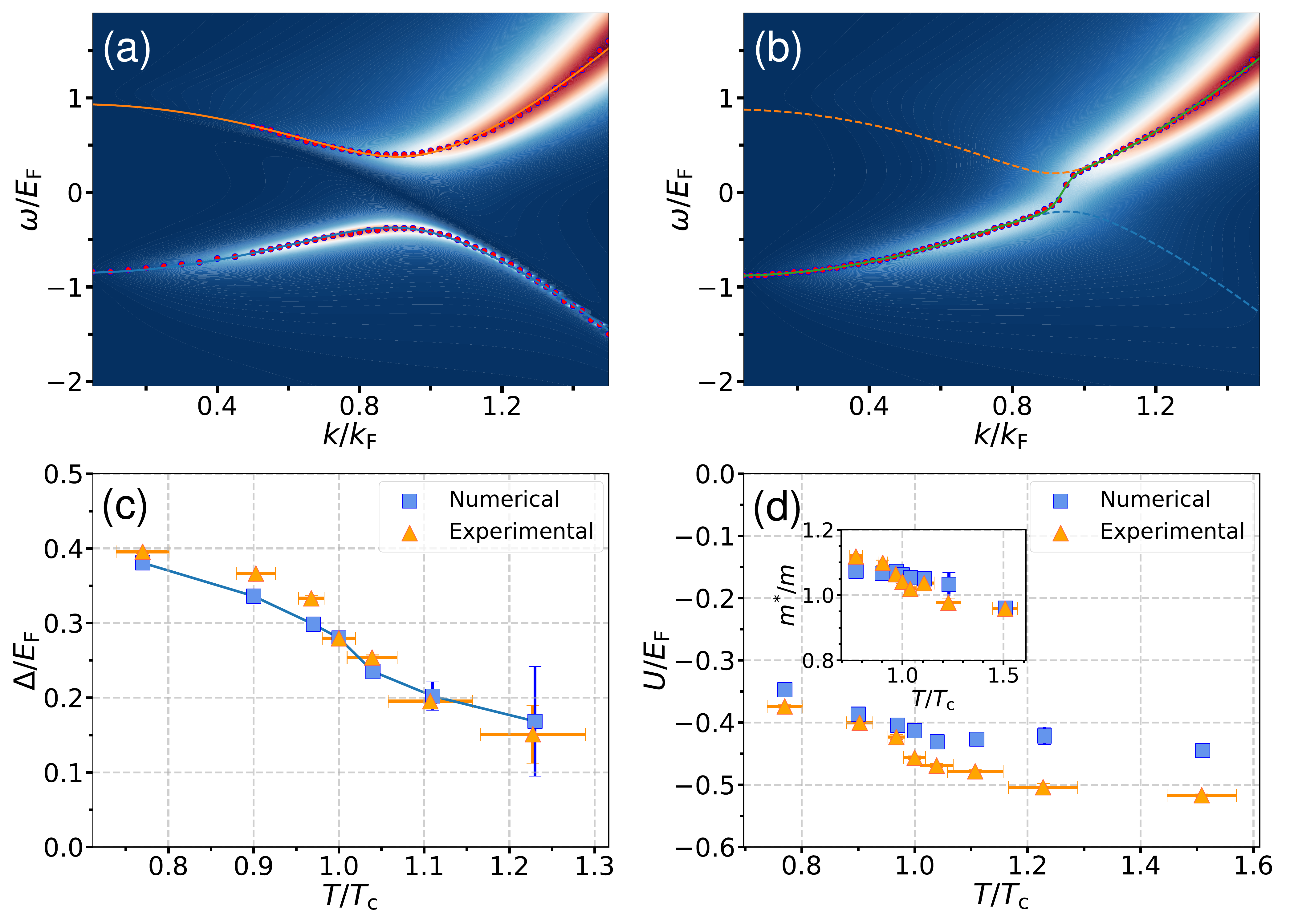}\vspace*{-1ex}
\caption{ Overlay of fitted dispersions on top of the spectral
  intensity map of $\mathbf{k}^{2}A(\mathbf{k},\omega)$ for (a)
  $T/T_\text{c}=0.9$ and (b) $1.11$.  Red dots indicate
  $E_\text{max}(k)$;  orange and blue lines correspond to
  $E_k^\pm$, respectively.  Comparison between theory (blue squares)
  and experiment (orange triangles) for (c) $\Delta$ and (d) $U$, as
  well as (inset) $m^*$, at different $T/T_\text{c}$.  Error bars
  represent one standard deviation.}
\label{fig:Akwfit2}
\end{figure}

To compare with experiment, we need to extract the gap from the
calculated spectral data, rather than using the initial input in
Eq.~(\ref{eq:iniAkw}), following the same experimental procedure of
data analysis as in Ref.~\cite{li2024nature}.  We fit with BCS-like
dispersions the locus of the maximum spectral response,
$E_\text{max}(k)$, in $\mathbf{k}^{2}A(\mathbf{k},\omega)$.  The
particle- and hole-like dispersions are given by
\begin{equation}
  \label{eq:BCSdispersion1}
  E_{{k}}^{(\pm)}=\pm\sqrt{\left(\frac{k^2}{2m^{*}}+U-\mu\right)^2+\Delta^2}\,,
\end{equation}
where $U$ is the chemical potential shift, and $m^{*}$ is the
renormalized fermion mass, both arising from the Hartree
energy.  The $S$-shaped dispersion is fitted with a weighted combination,
\begin{equation}
\label{eq:BCSdispersion2}
  E_{k}=a_{k}^{2}E_{k}^{(+)}+b_{k}^{2}E_{k}^{(-)}\,,
\end{equation}
where
$a_{k}^{2},b_{k}^{2}=\frac{1}{2}[1\pm\tanh(\frac{k-k_\text{c}}{\sigma})]$,
with $k_\text{c}$ and $\sigma$ representing the crossover momentum and
the crossover rate between the two branches, respectively.

In Fig.~\ref{fig:Akwfit2}, we present the fitted dispersion
$E_{{k}}^{(\pm)}$ (orange and blue lines) overlaid on top of the
spectral intensity map for (a) $T/T_\text{c}=0.9$ and (b) $1.11$ using
Eq.~(\ref{eq:BCSdispersion1}), along with an $S$-shape fit (green
line) in panel (b) using Eq.~(\ref{eq:BCSdispersion2}).  Panels (c)
and (d) show the fitting parameters $\Delta$, $U$, and $m^*$ (inset)
for the hole-like branch as a function of $T/T_\text{c}$.  We use
$E_k^{(-)}$ in Eq.~(\ref{eq:BCSdispersion1}) to fit the lower branch for
$0.77 \le T/T_\text{c} \le 1.04$, and $E_k$ in
Eq.~(\ref{eq:BCSdispersion2}) to fit S-shaped dispersions for $1.11
\le T/T_\text{c} < 1.51$.  For $T/T_\text{c}=1.51$, a parabolic fit
with $E^{}_{k}={k^2}/{2m^{*}}+U-\mu$ yields a lower error than the
S-shape fit.  The comparison of our theoretical values (blue squares)
with the corresponding experimental results (orange triangles) from
Ref.~\cite{li2024nature} shows a quantitative agreement, despite a
slight deviation in $U$ above $T_\text{c}$.  This deviation may have
to do with the fact that the numerical iteration is done only once. It
should also be noted that fitting the upper branch yields slightly
different results. Such particle-hole asymmetry becomes more
pronounced with increasing interaction strength beyond unitarity
\cite{Regular}.

\begin{figure}
\centering
\includegraphics[clip,width=3.2in]{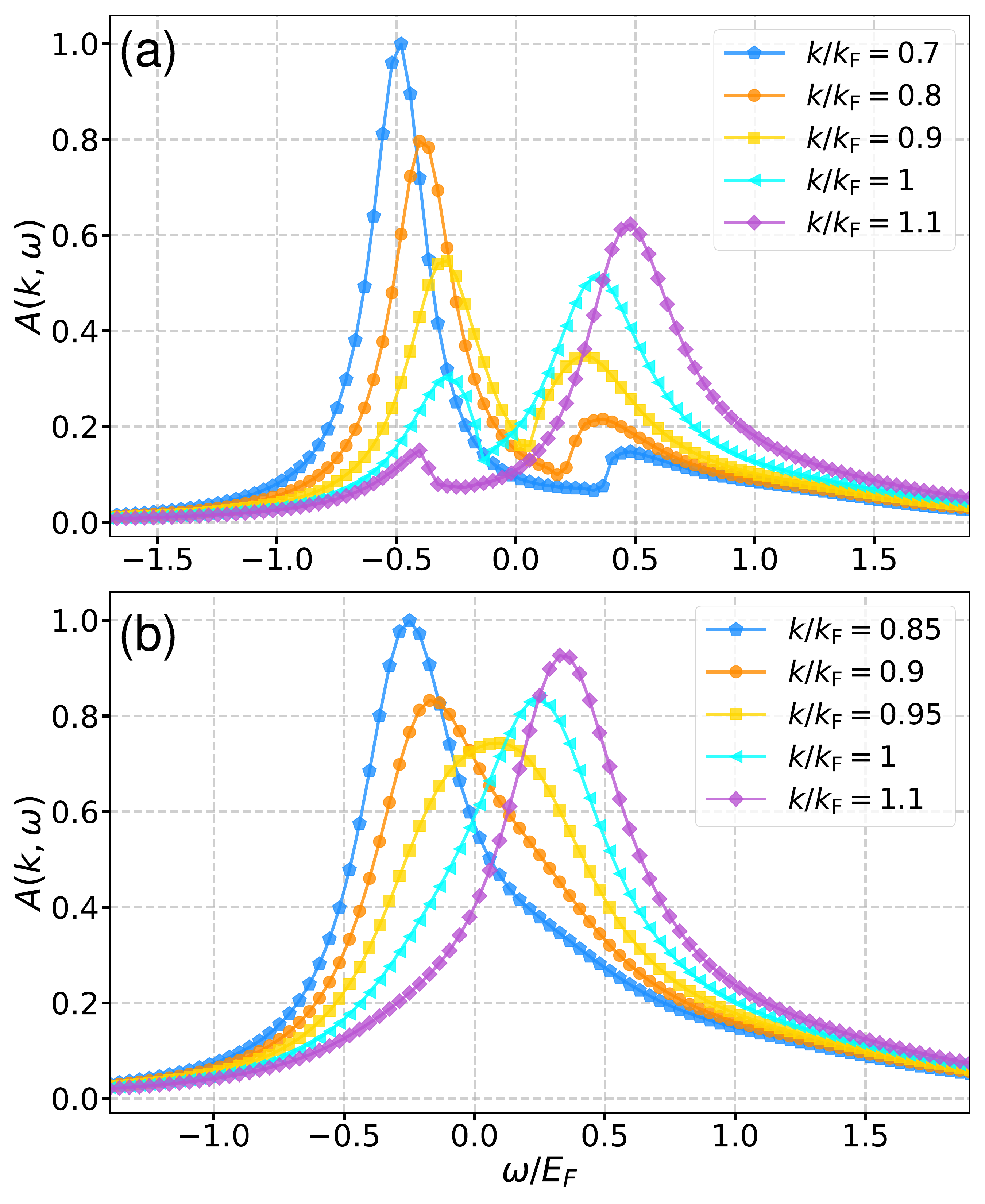}
\caption{ EDCs at (a) $T/T_\text{c}=1$ and (b) $T/T_\text{c}=1.11$ for
  various $k/k_\text{F}$ near $k_\mu$, showing the quasiparticle peak
  evolution. }
\label{fig:Akwfit3}
\end{figure} 

Next, shown in Fig.~\ref{fig:Akwfit3} is the evolution of the EDCs of
$A(\mathbf{k},\omega)$ for a series of $k/k_\text{F}$ not far from
$k_\mu$ at (a) $T/T_\text{c}=1$ and (b) $1.11$.  At $T_c$
(Fig.~\ref{fig:Akwfit3}(a)), the EDC is composed of two clearly
resolved spectral peaks, reflecting well-defined particle- and
hole-like quasiparticle energies.  As $k/k_\text{F}$ increases, the
spectral weight gradually shifts from the hole-like to the
particle-like branch, as described by the coherence factors
$v_\mathbf{k}^{2}$ and $u_\mathbf{k}^{2}$.  In contrast, for the above
$T_c$ case in panel (b), the EDCs exhibit only a single peak; here the
BCS-like quasiparticles are no longer sharply defined.  Nevertheless,
the strong asymmetry of the peaks and the large peak width at
$k/k_F=0.95$ do suggest that this single peak results from merging
two broad peaks rather than closing the pairing gap.

\begin{figure}
\centering
\includegraphics[clip,width=3.45in]{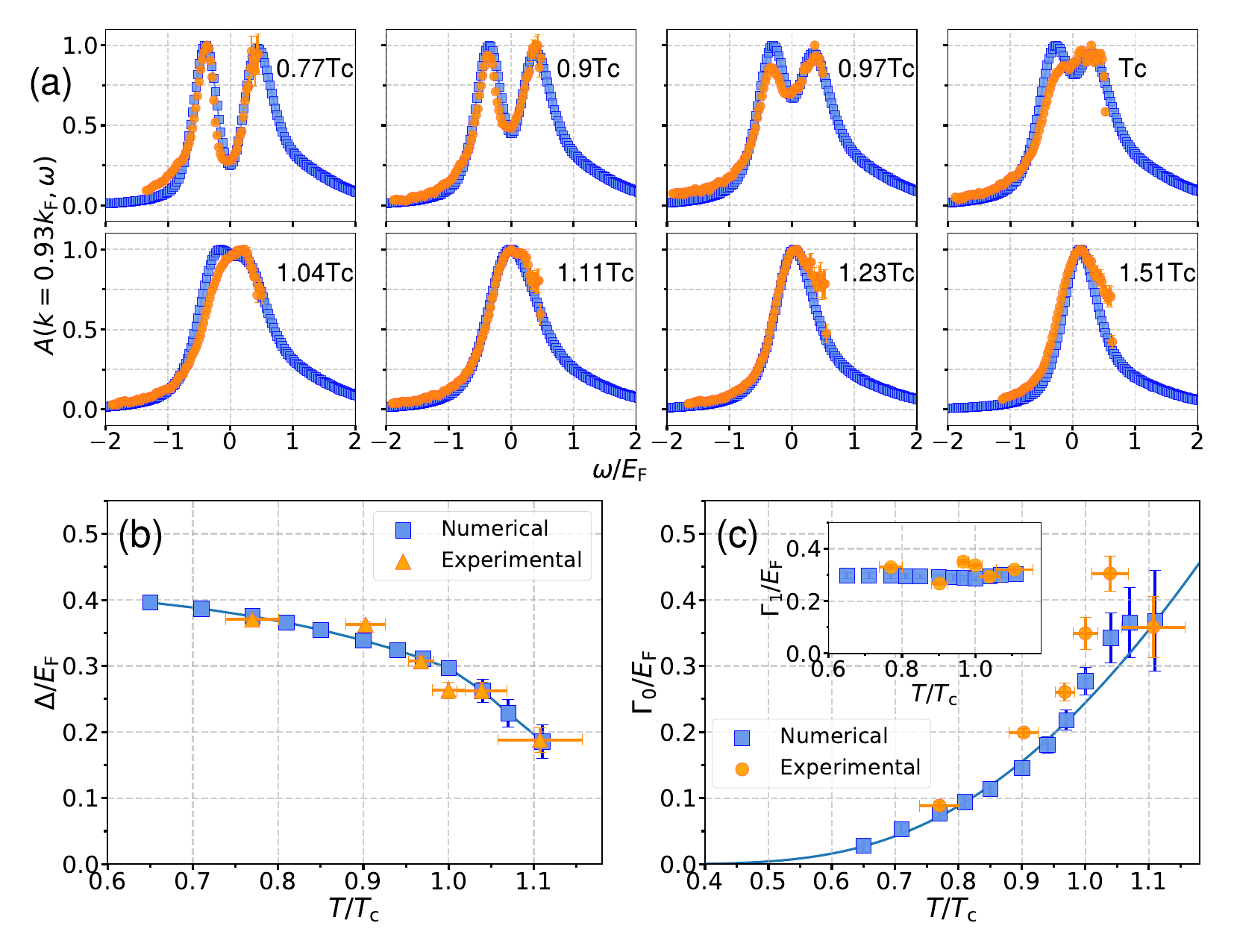}
\caption{ (a) Normalized EDCs of $A(\mathbf{k},\omega)$ from our
  calculations (blue) at $k=0.93k_\text{F}$ for different
  $T/T_\text{c}$, as labeled, (b) numerical $\Delta$ (blue) (c)
  $\Gamma_0$ and $\Gamma_1$ (inset) from the EDC fit. For comparison,
  the corresponding experimental data are also shown (in orange). The
  theoretical $\Gamma_0$ data are also fitted with an exponentially
  activated behavior (blue line) in (c).  Error bars represent one
  standard deviation. }
\label{fig:Akwfit4}
\end{figure}

Finally, we analyze the EDCs at the back-bending point
$k_\mu=0.93k_\text{F}$ (whose temperature dependence we ignore,
following the experiment~\cite{li2024nature}).  We also apply a
Gaussian broadening with a fixed standard deviation of
$0.16E_\text{F}$ when calculating the EDCs, to account for
instrumental resolution~\cite{li2024nature} which reflects the overall
error caused by the uncertainty in both frequency and momentum
measurements.  In Fig.~\ref{fig:Akwfit4}(a), we show our theoretical
normalized EDCs (blue squares) and compare with experimental data
(orange circles) at $k=0.93k_\text{F}$ for $T/T_\text{c}$ ranging from
$0.77$ to $1.51$.  With increasing $T$, the quasiparticle peaks
gradually broaden and merge, indicating a decreasing $\Delta$, and
thus a disappearing peak-to-peak separation.  The double peak
structure becomes a single broad peak as $T$ surpasses $T_\text{c}$,
even though the fitted $\Delta$ is still nonzero.  The calculated EDCs
match experimental data closely, with a slight discrepancy in the peak
height of the hole-like branch near $T_\text{c}$.  This primarily
arises from a slight deviation between the actual $k_\mu$ and the
assumed $k_\mu=0.93k_\text{F}$, which highlights the sensitivity of
EDCs to $k/k_F$ relative to the Fermi level, as can be seen in
Fig.~\ref{fig:Akwfit3}.

We fit the EDCs using the same phenomenological
self-energy model  \cite{Maly1997,Norman1998PRB,Chen2001} as in
Ref.~\cite{li2024nature}, given by
\begin{equation}
  \label{eq:selfenergy2}
    \Sigma(\mathbf{k},\omega)=\frac{\Delta^{2}}{\omega
      +\xi^*(\mathbf{k})+\mathrm{i}\Gamma_{0}}-{\mathrm{i}}\Gamma_{1}
    +[\xi^*(\mathbf{k})-\epsilon_\mathbf{k}+\mu]\,,
\end{equation}
where $\xi^*(\mathbf{k})\equiv \mathbf{k}^{2}/2m^{*}+U-\mu$ satisfying
$\xi^*(k_\mu)=0$, $\Gamma_0$ denotes the inverse pair lifetime, and
$\Gamma_{1}$ represents the $\omega$-independent single-particle
scattering rate. The expression of the spectral function derived from
this model via Eq.~(\ref{eq:Akw}) is used to fit the EDCs to extract
$\Delta$, $\Gamma_0$, and $\Gamma_1$. The results are shown in
Fig.~\ref{fig:Akwfit4}(b) and Fig.~\ref{fig:Akwfit4}(c), along with
the experimental data for quantitative comparison. The agreement is
good. In particular, we have a finite pseudogap at and above $T_c$,
with $\Delta \approx 0.3 E_\text{F}$ at $T_c$.
As shown in Fig.~\ref{fig:Akwfit4}(c), the theoretical data points of
$\Gamma_0$ (blue squares) fit nicely with an exponential, thermally
activated behavior, $\Gamma_{0}\propto \exp(-2\Delta_\text{0}/T)$
(blue curve), suggesting that the inverse pair lifetime is governed by
the virtual pair breaking and recombination process, with excitation
energy $2\Delta_{0}$. The fit yields $\Delta_{0}/E_\text{F}=0.41$,
slightly larger than the pairing gap at $T=0.65T_\text{c}$, the lowest
temperature accessed in experiment.  The single-particle scattering
rate $\Gamma_{1}$ shown in the inset of Fig.~\ref{fig:Akwfit4}(c)
remains nearly temperature independent, insensitive to pairing.
Performing Gaussian broadening on the EDCs mainly leads to an overall
increase in $\Gamma_{0}$ (and $\Gamma_{1}$) even at zero $T$, which is
removed by background subtraction.


In summary, we have performed a spectral study of the pseudogap
phenomena in unitary Fermi gases. By going beyond the previous pseudogap
approximation for the self-energy, and incorporating the particle-hole channel screening to
the interaction, we have arrived at a
quantitative agreement with the recent experiment \cite{li2024nature} on
the spectral function, and the associated pairing gap $\Delta$,
inverse pair lifetime $\Gamma_0$, and single-particle scattering rate
$\Gamma_1$, using the same data analysis procedure as in
Ref.~\cite{li2024nature}. In particular, the pseudogap self-energy is
calculated numerically using full convolution, which captures
explicitly the Hartree shift previously dropped (or hidden in chemical
potential) in the pseudogap approximation.
Our work thus provides a quantitative microscopic explanation for the
observed pseudogap and establishes pairing fluctuations as the key
mechanism underlying the pseudogap in unitary Fermi gases.


We thank Xi Li, Xing-Can Yao and Zhiqiang Wang for useful discussions.
This work was supported by the Quantum Science and Technology -
National Science and Technology Major Project (Grant
No. 2021ZD0301904).

\bibliography{References.bib}

\end{document}